\documentclass[RNAAS]{aastex631}

\begin{document}

\title{Broadening of the Main Sequence of a Star Cluster by Undetected Binaries}

\correspondingauthor{Steven Spangler}
\email{steven-spangler@uiowa.edu}

\author[0000-0002-4909-9684]{Steven R. Spangler}
\affiliation{Department of Physics and Astronomy, University of Iowa}

\begin{abstract}
This paper is concerned with the spread in apparent magnitudes (or absolute magnitudes) of main-sequence stars in a star cluster.  I specifically consider the effect of binary stars in broadening the main sequence.  I present analytic and semi-analytic expressions for the probability density function (pdf) of the magnitude at a given photometric color, including the effects of binarity.  The expressions obtained employ plausible models for the pdfs of the magnitudes of the primary and secondary stars, as well as the  distribution of secondary-to-primary mass ratio.  A crucial parameter is the fraction of stars in the star cluster that are binaries. The resultant formulas can be used to determine the fraction of binaries in a sample of stars taken from a star cluster, or to limit long term variations in the luminosity of solar type stars.
\end{abstract}

\section{1. Introduction} 
The Hertzsprung-Russell (HR) diagram of an open star cluster does not have a main sequence of infinitely narrow width, despite the fact that all stars are presumably of the same age and chemical composition. The causes for this broadening of the main sequence include photometric errors in the values of apparent magnitude and color, {\em possible} long term variations in luminosity of the star, and differential extinction across the cluster \citep{Curtis17}.  An additional, obvious contributor is binarity or multiplicity of the star system.  With the exception of obvious visual binary stars, the light from a binary will be blended and yield a single photometric value which is brighter than that of a single star of the same color. The incidence of binarity in field stars has been investigated by \cite{Raghavan10}, among others, and for open star clusters such as M67 \citep{Geller15, Geller21} and M35 \citep{Motherway24}. In these cases, the known binaries can be removed from the sample and the HR diagram replotted.  However, there will be residual, undetected binary and multiple systems that cause a corresponding residual broadening of the main sequence.   

This is the first in a planned set of three papers, with the goal of measuring or placing limits on long-term luminosity variations of solar-type stars.  The idea is that such information is contained in the broadening of the apparent magnitude distribution for stars with the same color in an open star cluster. Fuller elaboration of the concept will be given in a subsequent paper. This paper deals exclusively with the effect of undetected binary stars in an open cluster population in broadening the main sequence.  By undetected, I mean those stars that are binaries, but which have not been identified as visual, eclipsing, or spectroscopic binaries. I develop equations for the magnitude distribution (more precisely, the distribution of differences between the magnitude and a model for the main sequence), including a binary population.  Fitting these equations to an observed distribution could permit the binary contribution to be retrieved and corrected for, allowing measurements or limits to the intrinsic variations of the stars.

.\section{2. Basic Concepts}
I begin with some fundamental relations, which are used to derive the main formulas of the paper.  I start with basic definitions  \citep[e.g.][ Chapter 18]{Abell66}.  Let $F_1$ and $F_2$ be the fluxes (units such as Watts/m$^2$ contained in a bandpass such as the Johnson V filter) of the primary and secondary star.  Let $F_0$ be a ``fiducial flux'', which is most easily thought of as the flux of the primary alone, but need not be this precisely. Then the difference between the apparent (or absolute) magnitude of the binary stellar system and that of a system with the fiducial flux $F_0$ is
\begin{equation}
x \equiv \Delta m = 2.5 \log \left(\frac{F_1 + F_2}{F_0} \right)
\end{equation}
With this convention, $\Delta m > 0$ corresponds to a star brighter than $F_0$ (opposite of usual magnitude convention).  Now make the following assignments:
\begin{eqnarray}
F_1 = F_0y \\
 F_2=F_0 \left( \frac{M_2}{M_1}\right)^3 z = F_0 q^3 z 
\end{eqnarray}
where $q \equiv M_2/M_1$. The cubic dependence is a rough approximation to the mass-luminosity relation for the main sequence \citep[][p189]{Carroll07}.  Generalization to steeper forms is possible with the formalism presented in this paper.  

Equations (1) - (3) are obviously combined to form
\begin{equation}
x = 2.5 \log (y + q^3 z)
\end{equation}
The variables $y$ and $z$ represent the normalized fluxes of the primary and secondary stars.  I assume they have a mean value close to unity, and a relatively small variation about this value.  The variable $q$ is the mass ratio of the secondary to primary, and has a range 0 to 1. 

\section{3. Statistics of Variations}
The basic viewpoint of this paper is that for a given star system, $x$ will vary about a constant (close to zero) due to variations in $y$, $z$, and $q$, which are considered random variables, each described by it own probability density function (pdf), $f(y)$, $g(z)$, $\pi(q)$, respectively.  The variables $y$ and $z$ include all types of brightness variations of the stars, including intrinsic variations, noise in the measured variables, etc.  The mass ratio $q$ is also considered a random variable on the interval 0 to 1.  

For economy of notation and to simplify the derivation, let 
\begin{equation}
r \equiv y + q^3 z
\end{equation}
so $x = 2.5 \log (r)$.  The random variable $r$ is the sum of two parts, $y$ and $\theta \equiv q^3 z$.  
\subsection{ 3.1 Statistics of $\theta \equiv q^3z$}
The goal of the analysis in this section is to go from expressions for the pdfs of $q$ and $z$ to a pdf for $\theta \equiv q^3 z$.  This problem is discussed in \cite[][Section 2.4]{Beckmann67}. 
The approach is to transform from the original variables $(q,z)$ to new variables $(\theta(q,z),\tau(q,z))$. The variable $\theta$ is then the one of interest for which the pdf is sought, amd $\tau$ is a ``nuisance variable'' which is ultimately integrated over. I let $\theta$ be defined as above, and adopt as an arbitrary choice $\tau = q$.  Basic conservation of probability states that 
\begin{equation}
p_{\theta \tau}(\theta,\tau)d \theta d \tau = p_{qz}(q,z) dq dz
\end{equation}
This leads to \cite[][Section 2.4]{Beckmann67}
\begin{equation}
p_{\theta \tau}(\theta,\tau) = p_{qz}(q(\theta,\tau),z(\theta,\tau))\left| \frac{dq dz}{d\theta d\tau} \right|
\end{equation}
where the Jacobian of the transformation is given by 
\begin{equation}
\left| \frac{dq dz}{d\theta d\tau} \right| = \left| \left|\begin{array}{cr}
\frac{\partial q}{\partial \theta} & \frac{\partial q}{\partial \tau} \\ \frac{\partial z}{\partial \theta} & \frac{\partial z}{\partial \tau}   \end{array} \right| \right|
\end{equation}
that is, the absolute magnitude of the determinant of the $2 \times 2 $ matrix given on the right hand side of Equation (8).  

Given the definitions of $\theta(q,z)$ and $\tau(q,z)$ above, I form the inverses and have
\begin{eqnarray}
\frac{\partial q}{\partial \theta}=0, \frac{\partial q}{\partial \tau}=1 \nonumber\\
\frac{\partial z}{\partial \theta}=\frac{1}{\tau^3}, \frac{\partial z}{\partial \tau}=-\frac{3 \theta}{\tau^4}
\end{eqnarray}
Putting all of this together, I have
\begin{equation}
p_{\theta \tau}(\theta,\tau) = \frac{1}{\tau^3}\pi(\tau) g\left(\frac{\theta}{\tau^3}\right)
\end{equation}
Where I have used the physically-justified assumption of separability, 
\begin{equation}
p_{qz}(q,z) = \pi(q) g(z)
\end{equation}
The ``nuisance variable'' $\tau$ is integrated over to yield
\begin{equation}
p_{\theta}(\theta) = \int_0^1 d \tau p_{\theta \tau}(\theta,\tau) = \int_0^1 \frac{d \tau}{\tau^3} \pi(\tau) g\left(\frac{\theta}{\tau^3}\right) 
\end{equation}
which is the desired result. 
\subsubsection{ 3.1.1 Choice of Forms for pdfs}
To evaluate Equation (12), as well as prior expressions of interest, it is necessary to assume model forms for the density functions $f(y)$, $g(z)$, and $\pi(q)$.

For the flux variation functions $f(y)$ and $g(z)$ I choose a standard Gaussian, 
\begin{equation}
f(y) = \frac{1}{\sqrt{2 \pi} \sigma} \exp \left( -\frac{(y - \bar{y})^2}{2 \sigma^2}\right)
\end{equation}
For want of better information I choose the same function, with the same $\sigma$, for $g(z)$.  Once again, the functions $f(y)$ and $g(z)$ describe all processes that cause the measured flux of the star to differ from a constant value.  

The choice of the mass ratio distribution function $\pi(q)$ is less obvious, and relies on empirical input.  I interpret the results of \cite{Geller21} as indicating that, for binaries, the value of $q$ is {\em uniformly} distributed between 0 and 1 \citep[see Figure 12 of][]{Geller21}. This is a rough approximation to the results of \cite{Geller21}, but is used for its simplicity.  Single stars can be represented as binaries with $q=0$.  

Given the above, I adopt the following model expression for $\pi(q)$:
\begin{equation}
\pi(q) = (1-A) \delta(q) + A; \mbox{ for }   0 \leq q \leq 1
\end{equation}
This equation introduces the crucial parameter $A$, defined as the fraction of stars in the sample which are binaries. The first term in Equation (14) represents the single stars, and the second is the binaries, possessing secondaries with a uniform mass ratio distribution.   

With the results of Equations (12-14), I can calculate the first desired pdf, $p_{\theta}(\theta)$. This is given by
\begin{equation}
p_{\theta}(\theta)=A \int_0^1 \frac{d \tau}{\tau^3} g\left(\frac{\theta}{\tau^3}\right) 
\end{equation}
In addition to Equation (15) there is a delta function centered at $\theta=0$ with an amplitude of $(1-A)$. For ease in evaluating Equation (15), I change variables from $\theta \rightarrow \xi \equiv \frac{\theta}{\tau^3}$, so $\tau = \left(\frac{\theta}{\xi} \right)^{1/3}$.  The corresponding changes in the limits of integration are: $\tau = 0 \rightarrow \xi = \infty$ and $\tau = 1 \rightarrow \xi = \theta$. 

With this change of variables, Equation (15) becomes 
\begin{equation}
p_{\theta}(\theta)= \frac{A}{3 \theta^{2/3}} \int_{\theta}^{\infty} \frac{d \xi}{\xi^{1/3}} g(\xi) 
\end{equation}
Although not strictly necessary, the following algebra is simplified by approximating the Gaussian function $g(\xi)$ with an approximate function $g_1(\xi)$, represented by a boxcar function having the same area and second moment.
\begin{eqnarray}
g_1(\xi) = \frac{1}{2 \Delta}; \mbox{ if } 1-\Delta \leq \xi \leq 1+\Delta  \nonumber\\
= 0; \mbox{ if } \xi < 1-\Delta \mbox{ or } \xi > 1+\Delta
\end{eqnarray}
The second moments of $g(\xi)$ and $g_1(\xi)$ are the same if $\Delta = \sqrt{3} \sigma$, and I adopt that identity.  

I then have
\begin{equation}
p_{\theta 1}(\theta)= \frac{A}{3 \theta^{2/3}} \int_{\theta}^{\infty} \frac{d \xi}{\xi^{1/3}} g_1(\xi) 
\end{equation}
This expression readily yields three simple, algebraic expressions for $p_{\theta}$ in three ranges of $\theta$.  For $\theta < 1-\Delta$
\begin{eqnarray}
p_{\theta 1}(\theta)= \frac{A}{3 \theta^{2/3}} \left( \frac{1}{2 \Delta} \right) \int_{1-\Delta}^{1+\Delta} \frac{d \xi}{\xi^{1/3}}; \mbox{  if  } \theta < 1-\Delta \nonumber\\
= \frac{A}{3 \theta^{2/3}} \left( \frac{1}{2 \Delta} \right) \left( \left( \frac{3}{2} \right) [(1+\Delta)^{2/3} - (1-\Delta)^{2/3}] \right) \nonumber\\
= \frac{A}{3 \theta^{2/3}} \left( \frac{1}{2 \Delta} \right) C_1(\Delta)
\end{eqnarray}
For $1-\Delta \leq \theta \leq 1+\Delta$,
\begin{eqnarray}
p_{\theta 1}(\theta)= \frac{A}{3 \theta^{2/3}} \left( \frac{1}{2 \Delta} \right) \int_{\theta}^{1+\Delta} \frac{d \xi}{\xi^{1/3}};  \mbox{  if  } 1-\Delta \leq \theta \leq 1+\Delta \nonumber\\
= \frac{A}{3 \theta^{2/3}} \left( \frac{1}{2 \Delta} \right) \left( \left( \frac{3}{2} \right) [(1+\Delta)^{2/3} - \theta^{2/3}] \right) \nonumber\\
=\frac{A}{3 \theta^{2/3}} \left( \frac{1}{2 \Delta} \right) C_2(\Delta,\theta)
\end{eqnarray}
And for $\theta > 1+\Delta$,
\begin{equation}
p_{\theta 1}(\theta)=0; \mbox{ if } \theta > 1+\Delta
\end{equation}
The functions $C_1(\Delta)$ and $C_2(\Delta,\theta)$ are obviously defined by the terms in curved brackets in Equations (19) and (20), respectively.  

Equations (19) and (20) show that the dominant form of $p_{\theta }(\theta)$ or $p_{\theta 1}(\theta)$ is $\propto \theta^{-2/3}$. 
 A plot of $p_{\theta1}(\theta)$, henceforth taken to be equivalent to $p_{\theta }(\theta)$, is shown in Figure 1 for parameters $A=0.50$, $\sigma=0.12$. The binary fraction $A$, obviously influences the area under the curve, and thus the overall scaling, but not the shape.  The primary and secondary variability parameter $\sigma$ (and thus $\Delta$ ) determines the extent and shape of the ``rolloff'' to the function for $\theta \sim 1$.  
 
 \begin{figure}[h]
\begin{center}
\includegraphics[scale=0.50,angle=0]{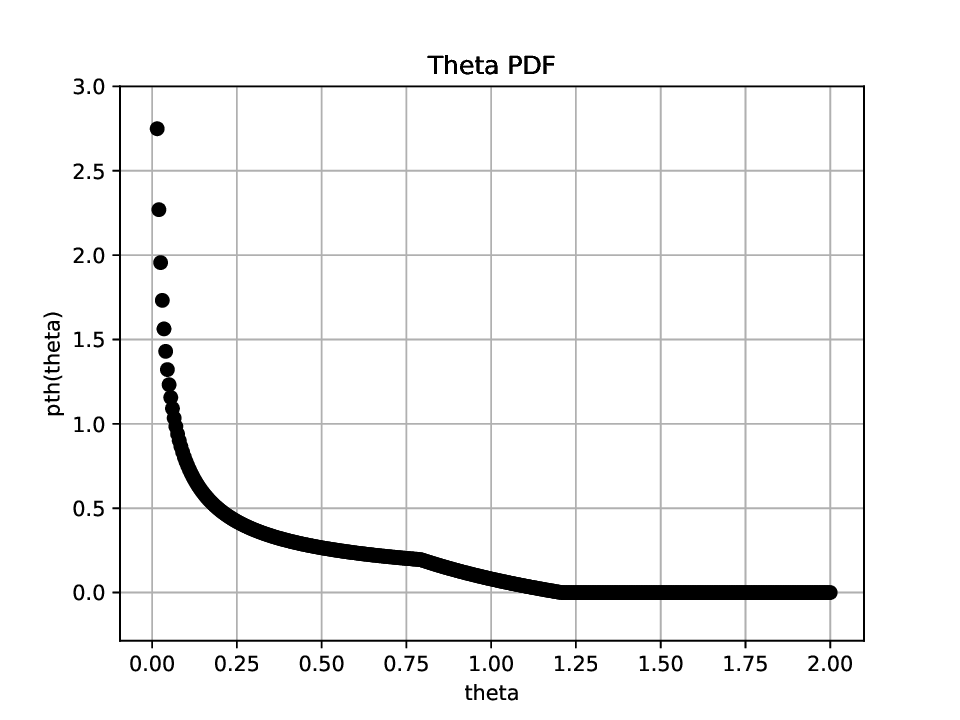}
\caption{The probability density function (pdf) of the flux contribution of the secondary star, $\theta$.  The plot shown is for the specific case $A =0.50$ (fraction of stars that are binary), and $\sigma = 0.12$ (rms fractional variation of both the primary and secondary).  The distribution shows a strong increase for small contributions to the total flux of the system. }
\end{center}
\end{figure}
As a reminder, it is worthwhile to recall that $p_{\theta }(\theta)$  represents the contribution of the secondary component to the overall flux of the star system. 
\subsection{3.2 Statistics of $r \equiv y + \theta$}
The pdf of the brightness variations (magnitude difference or Poynting flux differences) of the star system depends on the variations of the independent random variables $y$ and $\theta$, each of which is described by its own pdf.  

Equation (4) expresses the simple relationship that the net flux $r$ is the sum of $y$ and $\theta$, $r = y + \theta$.  We utilize the fact that the pdf of a quantity which is the sum of two random variables is the convolution of the pdfs of the variables in the sum \citep[][p76]{Beckmann67}.  In symbolic notation then, 
\begin{equation}
p_r(r)=p_y(y) \star p_{\theta} (\theta)
\end{equation}
where $p_y(y)$ is given by Equation (13) , and $p_{\theta} (\theta)$ is given by Equation (19), (20), or (21), depending on the value of $\theta$, and $\star$ indicates the operation of convolution.  

The first approach to this was done numerically, using the Python Numpy subroutine Convolve, with mode = ``full'',  to convolve two arrays containing the individual pdfs of $y$ and $\theta$. 

A difficulty which arose was that the $p_r(r)$ so derived was not properly normalized, in the sense that $\int_0^{\infty}p_r(r)dr \gg 1$ instead of equal to unity.  I conclude that this was due to the singularity in $p_{\theta}(\theta)$ that appears in the equations (19) - (21), or in the plot of Figure 1 \footnote{It should be noted that the singularity in $p_{\theta}(\theta=0)$ is not a problem, since analytically the integral $\int_0^{\infty} d \theta p_{\theta}(\theta)$ is unity.  The difficulty seems to arise when numerically convolving two arrays of numbers, one of which has a very large number in the first element.}. 

This problem was remedied by a procedure I call ``force normalization'', and is described as follows. 
\begin{enumerate}
\item The probability function $p_r(r)$ was calculated as described above, using the subroutine numpy.convolve. This distribution function was not normalized.
\item The area under the curve of this $p_r(r)$ was calculated numerically, and defined as $B_{norm}$.
\item All values of $p_r(r)$ were divided by $B_{norm}$ to generate a second estimate of $p_r(r)$ which is normalized to $A$. In addition, a Gaussian of the form of Equation (13), with an amplitude of $(1-A)$ was added to describe the single stars. This net, ``force-normalized'' function was adopted as the correct expression for $p_r(r)$. Its normalization was checked by the program. 
\end{enumerate}

A number of calculations with different numbers of points in the arrays representing the $y$ and $\theta$ pdfs were run.  Although the value of  $B_{norm}$ differed, the final resultant values of the force normalized $p_r(r)$ were essentially the same.  For this reason, I think the resultant expression for $p_r(r)$ used is accurate. An example of the function $p_r(r)$ calculated as above is shown in Figure 2. The asymmetric tail to the distribution for $r > 1$ is a clear indicator of the presence of binaries, and should be observationally accessible.   

\begin{figure}[h]
\begin{center}
\includegraphics[scale=0.50,angle=0]{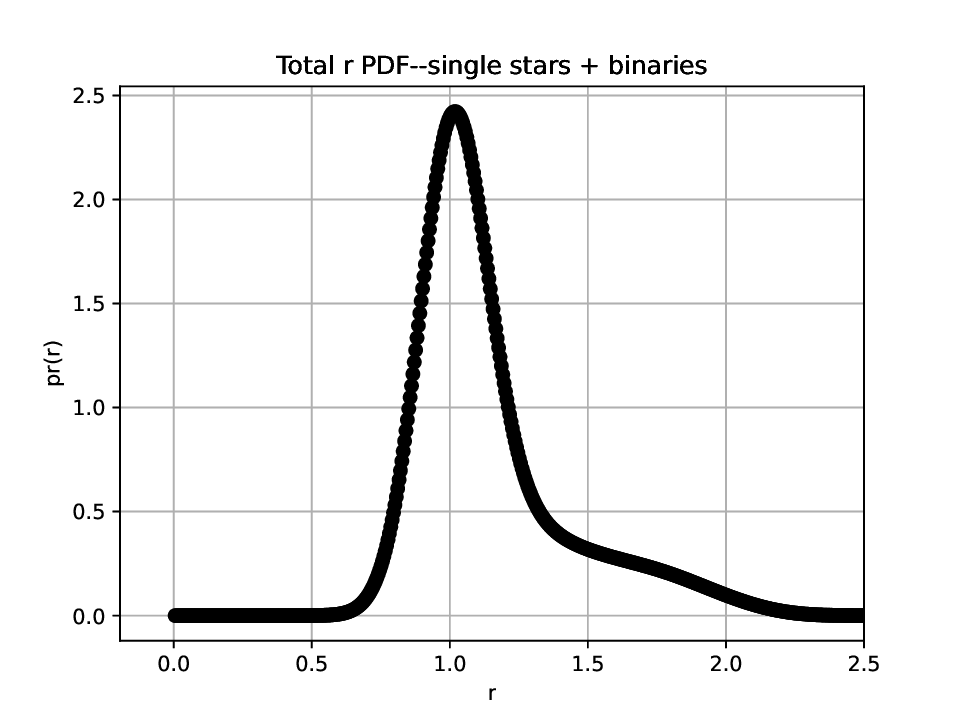}
\caption{The pdf of the normalized flux of both stars, r.  The plot shown is for the same case as Figure 1, i.e. $A =0.50$, and $\sigma = 0.12$.  The asymmetric tail for $r > 1$ should be an observable indicator of the incidence of binaries. }
\end{center}
\end{figure}
 
\section{4. Statistics of the Magnitude Difference $x \equiv \Delta m$}
Returning to Equation (4), one can obtain the statistics of $x = \Delta m$ from the statistics of $r$, given by $p_r(r)$.  

The equations relating $r$ and $x$ are 
\begin{equation}
x(r) = 2.5 \log (r) \mbox{;         }
r(x) = 10^{0.40x}
\end{equation}

From conservation of probability. 
\begin{equation}
p_r(r)dr = p_x(x) dx
\end{equation}
leading to 
\begin{equation}
p_x(x) = p_r(r(x)) \left| \frac{dr}{dx} \right|
\end{equation}
with the derivative at the right expressed as a function of $x$.
Use of relations (23) then yields
\begin{equation}
p_x(x) = p_r(r(x)) \left[0.4 \ln(10) \right] 10^{0.4x}
\end{equation}
Equation (26) allows $p_x(x)$ to be calculated directly from the $p_r(r)$ shown in Figure 2 (or any equivalent calculation).  In this calculation, uninterestingly large values of $x$ are simply discarded.  

A plot of the resultant, properly-normalized $p_x(x)$ is shown in Figure 3. This plot is directly relatable to the observational situation.  The ``plateau'' at $0.3 \leq x \leq 0.7$ is due to the (relatively large) binary contribution.  
\begin{figure}[h]
\begin{center}
\includegraphics[scale=0.50,angle=0]{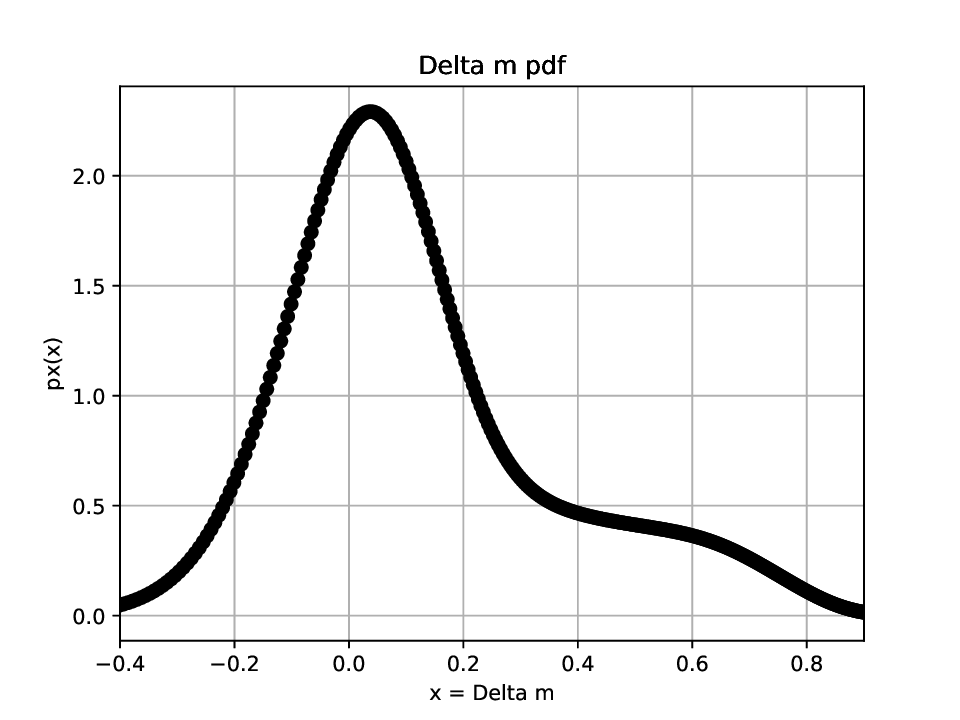}
\caption{The pdf of the magnitude difference $x \equiv \Delta m$. The magnitude difference is defined as the difference between the magnitude of a star system (single or binary) and the fiducial magnitude for a single star of the same color or effective temperature. The same convention is followed as throughout this paper, with $x > 0$ indicating a star brighter than the expectation value. The plot shown is for the same case as Figures 1 and 2, i.e. $A =0.50$, and $\sigma = 0.12$. An asymmetric ``bulge'' for $\Delta m > 0$ is a quantitative indicator of the binary population. }
\end{center}
\end{figure}
\section{5. Simulation of the $p_x(x)$ Distribution}
The approach in this paper is to employ relatively simple analytic expressions for all probability density functions, so that analytic methods may be used to the greatest extent feasible.  The advantage of this approach is the increased insight it provides.  However, numerical simulations are very important to verify the expressions and plots shown to this point, and to insure that conceptual or mathematical blunders have not been made.  

I undertook a numerical simulation of the pdf $p_x(x)$, adopting the the individual pdfs of $y$, $q$, and $z$ given in Equations (13) and (14).  The following steps were taken. 
\begin{enumerate}
\item A number was chosen for $N_{\star}$, the number of simulated stars. 
\item For each simulated star, a normalized flux for the primary was chosen according to Equation (13), with $\bar{y}=1$ and $\sigma$  selectable.  The values were chosen using the Python subroutine numpy.random.normal.  
\item The secondary stars, describable by the binary fraction $A$, were input to the simulation in the following way.  For the first $AN_{\star}$ elements in the array containing the normalized secondary flux $\theta$, a value was chosen in the way described in item 4 below.  For the remaining $(1-A)N_{\star}$ elements, $\theta$ was set to zero.
\item For the simulated binaries (i.e. $\theta \neq 0$), $\theta$ was calculated as $\theta = q^3z$, with $z$ also selected as Gaussian-distributed random variables in the same way as described in item 2, and with the same value of $\sigma$.  The value of $q$ was chosen from a uniform random distribution using the Python subroutine numpy.random.uniform.  
\item For each simulated star, the values of $y$ and $\theta$ were summed to form $r$, and the corresponding value of $x$ chosen using Equation (4) or (23).  All of the simulated $x$ values were binned into a histogram, the bin width of which can be chosen.  
\item The theoretical $p_x(x)$ was calculated, using Equations (22) and (26), for the same value of $A$ and $\sigma$ as the simulation.  It was converted to a prediction of the number of stars per bin, using the appropriate values of $N_{\star}$ and bin size, and overplotted on the simulation results.  A comparison of the two then provides a test of the validity of the theoretical expression. The resultant plot is considered the ``right answer''. 
\end{enumerate}
An advantage of the simulation is that it does not make any of the assumptions in the derivations of Sections 2 - 4 (other than that of the fundamental pdfs of $y$, $q$, and $z$), and does not require the ``forced normalization'' operation described in Section 3.2. 

A comparison of my semi-analytic expression for $p_x(x)$ and the purely simulation result is shown in Figure 4.  I consider the agreement entirely satisfactory. Similar results have been obtained for various values of $A$ and $\sigma$.  
\begin{figure}[h]
\begin{center}
\includegraphics[scale=0.50,angle=0]{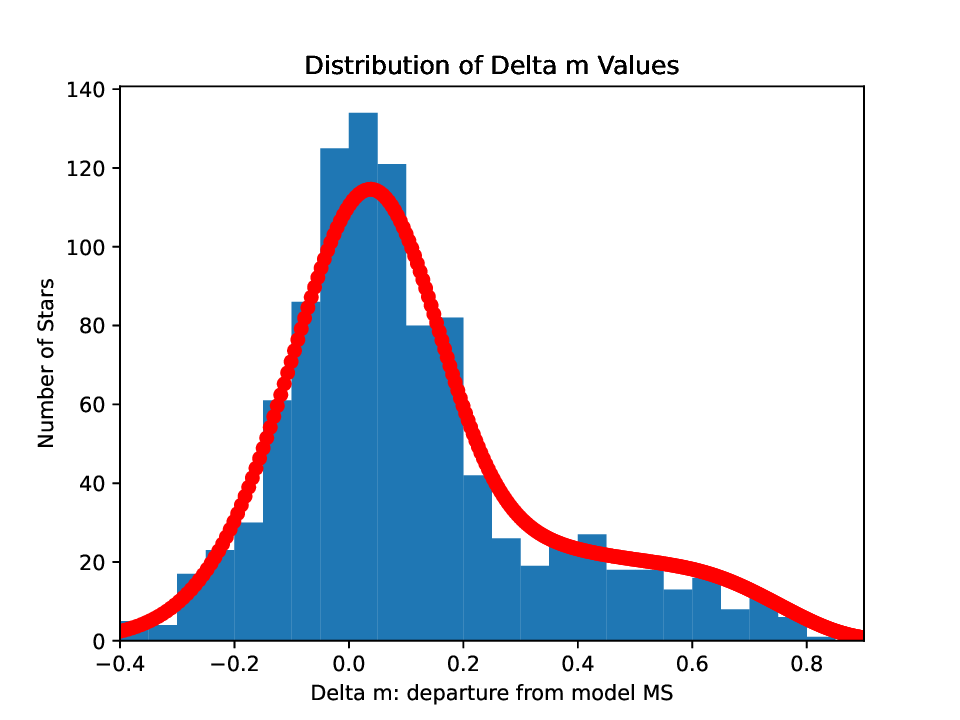}
\caption{Results of a simulation of a sample of 1000 stars with $A=0.50$ binarity fraction and a variability parameter $\sigma=0.12$.  The results of the simulation have been grouped in bins of $\Delta m = 0.05$.  The heavy red line is the theoretical expression derived in sections up through Section 4, for the same parameters.  The agreement is clearly satisfactory, indicating the validity of our expression for the pdf $p_x(x)$.}
\end{center}
\end{figure}
\section{6. A Fully Analytic Approximation for $p_r(r)$}
The discussion to this point may be termed ``semi-analytic'' because it relied on a numerical approach to convolve the pdfs of $y$ and $\theta$ (i.e. Equation (22)). This numerical approach was done because of the ease in doing so, and because of the singularity in $p_{\theta}(\theta)$. In this section, I explore the possibility that, with a slight additional approximation to $p_{\theta}(\theta)$, a fully analytic expression may be obtained for $p_r(r)$. Once this is in hand, the magnitude difference pdf $p_x(x)$ can be obtained in straightforward fashion using Equation (26). 

In explicit form, Equation (22) is written as
\begin{eqnarray}
p_r(r) = \int d \theta p_{\theta} (\theta) p_y(r - \theta) \nonumber\\
 = \int dy p_y(y) p_{\theta} (r-y)
\end{eqnarray}
In the calculation that follows, I choose the first form of the convolution in Equation (27).  The calculation is slightly more convenient if the $p_y(y)$ pdf is shifted to being centered at the origin rather than unity.  This shift is added back in at the end of the calculation.  

I adopt an approximation in carrying out the convolution, Equation (27).  The pdf $p_{\theta}(\theta)$ is approximated by Equation (19) for $\theta \leq 1$, and 0 for $\theta > 1$.  This obliterates the rolloff due to variability of the secondary star. This approximation will be better the smaller $\sigma$ and $\Delta$. 

Use of this approximation gives a corresponding approximation for  $p_r(r)$, which I term $p_{r2}(r)$, 
\begin{equation}
p_r(r) \simeq  p_{r2}(r)= \frac{B}{\sqrt{2 \pi} \sigma} \int_{0}^{1} d \theta\left( \frac{1}{\theta^{2/3}} \right) \exp \left( -\frac{(r-\theta)^2}{2 \sigma^2} \right)
\end{equation}
where $B \equiv \frac{A C_1(\Delta)}{6 \Delta}$ with $A$, $C_1(\Delta)$, and $\Delta$ defined above. 

This integral can be approached by making a change of variable, $\theta \rightarrow t \equiv \frac{\theta}{\sqrt{2}\sigma}$.  With this change of variables, Equation (28) becomes 
\begin{equation}
p_{r2}(r)= \frac{B}{\sqrt{\pi}} \left( \frac{1}{2^{1/3}\sigma^{2/3}} \right) \int_{0}^{t_1} dt \left( \frac{1}{t^{2/3}} \right) \exp \left( -(t_0-t)^2 \right)
\end{equation} 
with $ t_0 \equiv \frac{r}{\sqrt{2} \sigma}$ and $ t_1 \equiv \frac{1}{\sqrt{2} \sigma}$.
This equation can be written more compactly as 
\begin{eqnarray}
p_{r2}(r)= \frac{B}{\sqrt{\pi}} \left( \frac{1}{2^{1/3}\sigma^{2/3}} \right) I_2(t_0), \nonumber\\
I_2(t_0) \equiv \int_{0}^{t_1} dt \left( \frac{1}{t^{2/3}} \right) \exp \left( -(t_0-t)^2 \right)
\end{eqnarray} 
With $I_2(t_0)$ now a specialized function defined by the relatively simple integral in Equation (30). Since $I_2(t_0)$ is a function of the single variable $t_0$ (once the upper limit of integration, $t_1$ is specified), it could be tabulated and used without further computation. The $r$ dependence is now carried by the equivalent variable $t_0$.

It should be pointed out that in applications of interest, $\sigma \ll 1$, so that $t_1 \gg 1$, and $I_2(t_0)$ may be replaced by a simpler and more aesthetic expression in which the upper limit of integration is replaced by $\infty$. 

The function $p_{r2}(r)$ is trivial to calculate, apart from the specialized function $I_2(t_0)$.  For a range of interest of $r$, $I_2(t_0)$ was calculated by a Mathematica notebook which computed $I_2(t_0)$ using the subroutine NIntegrate.  The corresponding value of $p_{r2}(r)$ was then computed, and compared with the standard, semi-analytic expression derived as described in Section 3.2.  A comparison of the two expressions is given in Figure 5.  
\begin{figure}[h]
\begin{center}
\includegraphics[scale=0.50,angle=0]{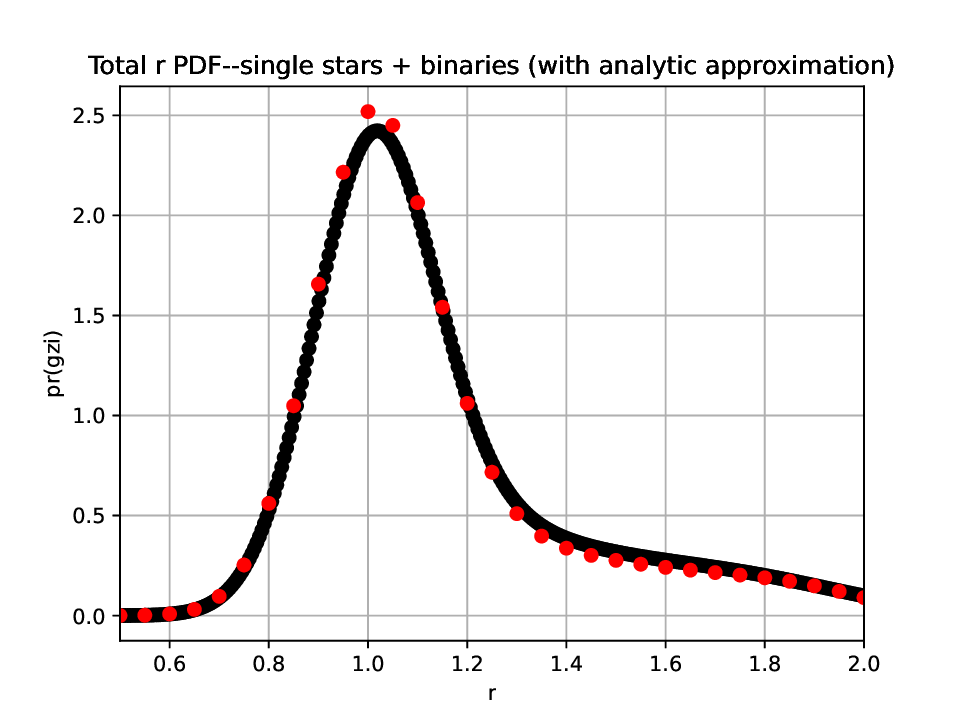}
\caption{Comparison of results from a partially-numerical calculation of $p_r(r)$ (as described through Section 4, including use of numerical convolution of the $p_y(y)$ and $p_{\theta}(\theta)$ distributions, solid black line), with a ``completely analytic'' approximation (red dots).  The agreement is clearly excellent, indicating that Equation (30) could be used for $p_r(r)$, with use of Equation (26) to convert it to $p_x(x)$.  }
\end{center}
\end{figure}
The agreement between the two expressions is obviously completely satisfactory, and indicates that Equation (30) could be used as the basis for an analytic calculation of $p_x(x)$, with Equation (26) (a straightforward, algebraic relation) for the conversion between $p_r(r)$ and $p_x(x)$.  I have not done this here because the agreement shown in Figure 5 demonstrates that Equation (30) is an adequate approximation for $p_r(r)$, which is where the computational challenges are, and the transformation from $p_r(r)$ to $p_x(x)$ is straightforward and discussed in Section 4.  
\section{7. The $\Delta m$ Distribution for a ``Culled'' Stellar Sample}
The goal of this research program is to quantitatively determine the effect of a binary population on the distribution of apparent (or absolute) magnitudes of main-sequence stars with the same effective temperature or color. With the mathematical models developed here, it would be possible to retrieve the parameters $A$ and $\sigma$ from an observed distribution. 

For purposes of illustration, in the cases presented above I have chosen distributions with a relatively large binary fraction $A=0.50$.  This is a reasonable approximation for a sample of field stars in the solar vicinity \citep{Raghavan10}. The asymmetric tail due to binaries in Figures 2-5 is then particularly evident.  These cases emphasize that the formalism developed here could be used to retrieve the fraction of binaries, $A$, even in the absence of spectroscopic or photometric information that would reveal the presence of spectroscopic or eclipsing binaries.  

An application of greater interest to me is constraining the variations of the primary star, given by the parameter $\sigma$.  As briefly noted in the Introduction, there are many effects that contribute to $\sigma$, but one of great interest is long term luminosity variations of main-sequence stars similar to the Sun, which would broaden the distribution of primary fluxes, and thus $x$.  

In this case, best results would be obtained when the number of binaries has been reduced to the smallest value possible.  The effect of binaries could then be considered a perturbation to a distribution which is dominated by dispersion in the brightness of the primary.  The parameterization of the primary variations and the properties of the secondaries used in this paper employ approximations, some of which are brutal.  The smaller the number of binaries actually present, the less consequential the approximation of the secondary properties, given by Equations (3) and (14) will be. I use the term ``culled'' sample to indicate a stellar sample in which a significant number of binaries have been identified and can be removed, leaving a residual (culled) sample for which $A$ is reduced.  

This situation can actually be realized.  \cite{Geller15} and \cite{Geller21} have identified and determined the orbital elements for a significant portion of the binaries in the open cluster M67, and \cite{Motherway24} have done the same for binaries in M35.  Thus, a culled stellar sample is available for both of these open star clusters.  
\begin{figure}[h]
\begin{center}
\includegraphics[scale=0.50,angle=0]{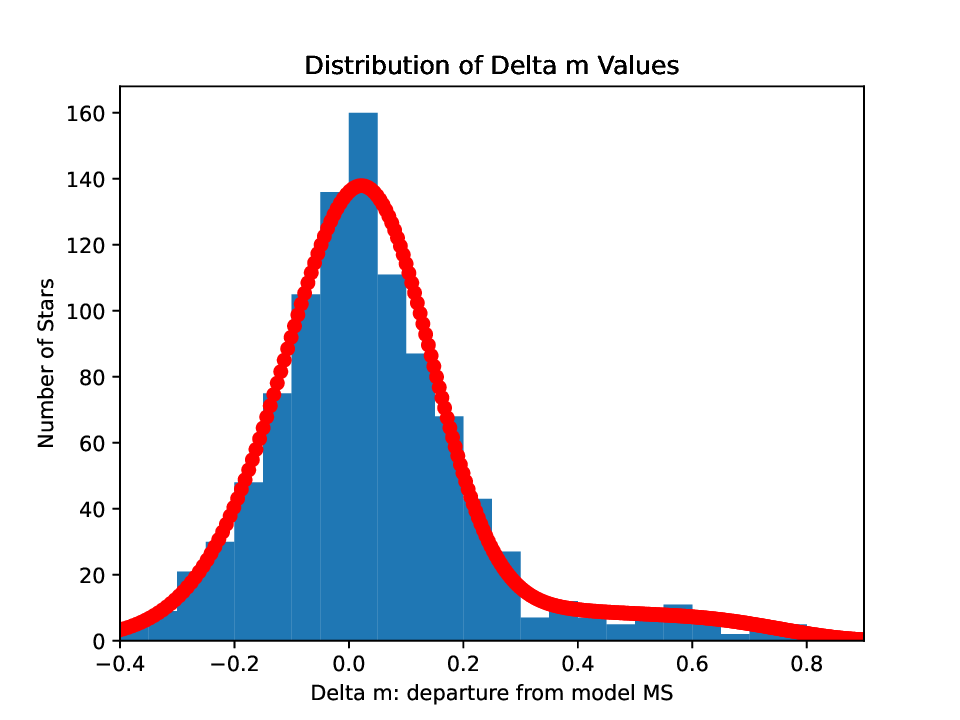}
\caption{ A distribution of stars for which the binaries have been partially ``culled'' or edited by independent information.  The parameters of the distribution are $\sigma=0.12$, $A=0.20$. Same format as Figure 4, with a comparison to a simulation.  For this case, the peak value of $p_x(x)$ is 2.75 at a value of $x=0.020$. The asymmetric tail due to binaries is clearly reduced in this case.}
\end{center}
\end{figure} 
To illustrate what the $x \equiv \Delta m$ distribution for such a culled sample might look like, Figure 6 shows the calculated $p_x(x)$ function, together with a Monte Carlo simulation like that presented in Section 5, for a distribution in which $A = 0.20$.  The properly normalized $p_x(x)$ function can be retrieved from this plot by noting that the peak value is $p_x(\mbox{xmax})=2.75$ for xmax=0.02. Comparison of Figure 6 with Figure 4 shows that for a culled sample, the effect of the unremoved binaries is less pronounced, and an approximate scheme for their removal should be more successful.  
\section{8. Summary and Conclusions}
\begin{enumerate}
\item This paper is concerned with the dispersion of apparent (or absolute) magnitudes of main sequence stars about a model for the color-magnitude relation for a stellar sample.  The most obvious and interesting example is that of an open star cluster.  This dispersion is due to two processes; variations in the measured magnitude of the primary and secondary (if present) stars, and the presence of binary companions to stars which obviously brighten the composite star.  

\item I have obtained two simple expressions for the probability density function (pdf) for the apparent or absolute magnitudes of such a population of stars.  One expression is semi-analytic, another is completely analytic, but is more approximate.  The semi-analytic expression has been confirmed by Monte Carlo simulations, including a dispersion in the luminosities of the primary and secondary stars, as well as a range of secondary masses. In either case, the pdf is determined by two parameters, $A$, the fraction of stars in the sample that are binaries, and $\sigma$, the percentage variation of the brightness of the primary (and secondary) due to measurement error, different distances, and intrinsic variability. 

\item The resultant expressions could be of use in two types of investigation.  In one, the fraction of binary stars in a star cluster or other stellar sample could be determined solely from photometry observations at a single epoch, without the need for numerous spectroscopic or photometric observations to detect spectroscopic and eclipsing binaries, respectively. The other application is to estimate the contribution of undetected binaries to  the broadening of the main sequence of a star cluster, so that effect can be removed and a better estimate is possible of any long-term intrinsic luminosity variations of the primary stars.  
\end{enumerate}

\end{document}